\documentclass[12pt]{article}
\usepackage{amsfonts,amssymb,amsmath}
\usepackage{epsfig}
\makeatletter
\@addtoreset{equation}{section}

\makeatother

\newcommand\RR{\mathbb R}
\newcommand\CC{\mathbb C}
\newcommand\cA{{\cal A}}
\newcommand\cB{{\cal B}}
\newcommand\cC{{\cal C}}
\renewcommand{\Re}{\mathop{\mathrm{Re}}}
\renewcommand{\Im}{\mathop{\mathrm{Im}}}

\newcommand\beq{\begin{equation}}
\newcommand\eeq{\end{equation}}

\newtheorem{theorem}{Theorem}

\newtheorem{remark}{Remark}
\newtheorem{proposition}{Proposition}
\newtheorem{statement}{Statement}

\makeatletter
\@addtoreset{theorem}{section}
\@addtoreset{lemma}{section}
\@addtoreset{definition}{section}
\@addtoreset{remark}{section}
\@addtoreset{proposition}{section}
\@addtoreset{statement}{section}
\makeatother

\begin{document}
\title{Faddeev eigenfunctions for multipoint potentials
\thanks{The main part of the work was fulfilled during the visit of the 
first author to the Centre de Math\'ematiques Appliqu\'ees of \'Ecole Polytechnique
in October 2012. The work was also partially supported by the 
Russian Federation Government grant  No~2010-220-01-077.
The first author was also partially supported by 
Russian Foundation for Basic Research grant 11-01-12067-ofi-m-2011, 
and by the program  ``Fundamental problems of nonlinear dynamics'' of the 
Presidium of RAS. The second author was also partially supported by FCP Kadry No. 14.A18.21.0866.}}
\author{P.G. Grinevich
\thanks{Landau Institute of Theoretical Physics, Kosygin street 2, 117940 Moscow, Russia;
Moscow State University, Moscow, Russia; Moscow Physical-Technical Institute, 
Dolgoprudny, Russia; e-mail: pgg@landau.ac.ru} 
\and R.G. Novikov\thanks
{CNRS (UMR 7641), Centre de Math\'ematiques Appliqu\'ees, 
\'Ecole Polytechnique, 91128, Palaiseau, France; 
e-mail: novikov@cmap.polytechnique.fr}}
\date{}
\maketitle
\begin{abstract}
We present explicit formulas for the Faddeev eigenfunctions and related 
generalized scattering data for multipoint potentials in two and three
dimensions. For single point potentials in 3D such formulas were obtained in 
an old unpublished work of L.D. Faddeev. For single point potentials in 2D such 
formulas were given recently in \cite{GRN}.
\end{abstract}

\section{Introduction}
Consider the Schr\"odinger equation 
\beq
\label{eq:1}
-\Delta\psi + v(x)\psi = E\psi,  \ \ x\in\RR^d, \ \ d=2,3,
\eeq
where $v(x)$ is a real-valued sufficiently regular function on $\RR^d$ with 
sufficient decay at infinity. 

Let us recall that the classical scattering eigenfunctions $\psi^+$ for 
(\ref{eq:1}) are specified by the following asymptotics as $|x|\rightarrow\infty$:
\begin{align}
\label{eq:3a}
&\psi^+= e^{ikx}-i\pi\sqrt{2\pi}e^{-\frac{i\pi}{4}}f\left(k,|k|\frac{x}{|x|} 
\right)
\frac{e^{i|k||x|}}{\sqrt{|k||x|}} +o\left(\frac{1}{\sqrt{|x|}} \right), \ \ &
d=2,\\
\label{eq:3b}
&\psi^+= e^{ikx}-2\pi^2 f\left(k,|k|\frac{x}{|x|}\right)
\frac{e^{i|k||x|}}{|x|} +o\left(\frac{1}{|x|} \right), \ \ &
d=3,
\end{align}
$x\in\RR^d$, $k\in\RR^d$, $k^2=E>0$, where a priori unknown function 
$f(k,l)$, $k,l\in\RR^d$, $k^2=l^2=E$, arising in (\ref{eq:3a}), (\ref{eq:3b}), is 
the classical scattering amplitude for (\ref{eq:1}). In addition, we 
consider the Faddeev eigenfunctions $\psi$ for (\ref{eq:1}) specified by
\beq
\label{eq:4}
\psi= e^{ikx}\left(1+o(1) \right) \ \
\mbox{as} \ \ |x|\rightarrow\infty, 
\eeq
$x\in\RR^d$, $k\in\CC^d$, $\Im k\ne 0$, $k^2=k_1^2+\ldots+k_d^2=E$;
see \cite{F1}, \cite{N0}, \cite{G}. The generalized 
scattering data arise in more precise version of the expansion (\ref{eq:4}) 
(see also formulas (\ref{eq:18})-(\ref{eq:21})). The Faddeev eigenfunctions have 
very rich analytical properties and are quite important for inverse scattering 
(see, for example, \cite{F2}, \cite{NKh}, \cite{G}).

In the present article we consider equation~(\ref{eq:1}), where $v(x)$ is
a finite sum of point potentials in two or three dimensions 
(see \cite{BF}, \cite{AGHH} and references therein). We will write these 
potentials as: 
\beq
\label{eq:2}
v(x)=\sum\limits_{j=1}^{n} \varepsilon_j\delta(x-z_j),
\eeq
but the precise sense of these potentials will be specified below 
(see Section~\ref{sect:3}) and, strictly speaking, $\delta(x)$ is not 
the standard Dirac delta-function (in the physical literature the term
renormalized $\delta$-function is used). 

It is known that for these multipoint potentials the classical scattering 
eigenfunctions $\psi^+$ and the related scattering amplitude $f$ can be
naturally defined and can be given by explicit formulas (see \cite{AGHH} and 
references therein). In addition, for single point potentials explicit
formulas for the Faddeev eigenfunctions $\psi$ and related generalized scattering
amplitude $h$ were obtained in an old unpublished work by L.D. Faddeev for 
$d=3$ and in \cite{GRN} for $d=2$. 

In the present article we give explicit formulas for the Faddeev functions 
$\psi$ and $h$ for multipoint potentials in the general case for real energies
in two and three dimensions (see Theorem~\ref{th:3.1}  from the 
Section~\ref{sect:3}). Let us point out that our formulas for 
$\psi$ and $h$ involve the values of the  Faddeev Green function $G$ for 
the Helmholtz equation, where 
\begin{align}
\label{eq:1.6}
&G(x,k)=-\frac{1}{(2\pi)^d} \ e^{i k x}  \int\limits_{\RR^d} 
\frac{e^{i\xi x}}{\xi^2+2k\xi}  d\xi,  \\
&(\Delta + k^2) G(x,k) = \delta(x), \ \ \ \ x\in\RR^d, \ \  k\in\CC^d, \ \ \Im k\ne 0.
\end{align}
In the present article we consider $G(x,k)$ as some known special function.

In addition, basic formulas and equations of monochromatic inverse scattering, 
derived for sufficiently regular potentials $v$, remain valid for the Faddeev functions 
$\psi$ and $h$ of Theorem~\ref{th:3.1}. Thus, basic formulas and equations 
of monochromatic inverse scattering are illustrated by explicit examples related to 
multipoint potentials. We think that the results of the present work can be used, 
in particular, for testing different monochromatic inverse scattering algorithms 
based on properties of the Faddeev functions $\psi$ and $h$ (see \cite{BBMR}
as a work in this direction).

It it interesting to note also that explicit formulas for $\psi$ and $h$ for multipoint
potentials show new qualitative effects in comparison with the one-point case. 
In particular, the Faddeev eigenfunctions for 2-point potentials in 3D may have singularities 
for real momenta $k$, in contrast with the one-point potentials in 3D 
(see Statement~\ref{st:3.1}).   

Besides, functions  $\psi$ and $h$ of Theorem~\ref{th:3.1} for $d=2$ illustrate a very 
rich family
of 2D potentials with spectral singularities in the complex domain. Let us recall that 
monochromatic 2D inverse scattering is well-developed only under the assumption 
that such singularities are absent at fixed energy (see \cite{GN}and \cite{GRN} 
for additional discussion in this connection). We hope that the aforementioned  examples 
and quite different examples from  \cite{G0}, \cite{TT}
will help to find correct analytic formulation of monochromatic inverse scattering 
in two dimensions in the presence of spectral singularities.

\section{Some preliminaries}

It is convenient to write 
\beq
\label{eq:5}
\psi= e^{ikx}\mu,
\eeq
where $\psi$ solves (\ref{eq:1}), (\ref{eq:4}) and $\mu$ solves
\beq
\label{eq:6}
-\Delta\mu -2i k\nabla\mu+ v(x) \mu=0, \ \ \ \ k\in\CC^d, \ \ 
k^2=E.
\eeq

In addition, to relate eigenfunctions and scattering data it is 
convenient to use the following presentations, used, for example, 
in \cite{N2} for regular potentials:

\beq
\label{eq:18}
\mu^+(x,k)=1 -\int\limits_{\RR^d} \frac{e^{i\xi x}F(k,-\xi)}
{\xi^2+2(k+i0k)\xi } d\xi, \ \ k\in\RR^d\backslash0,
\eeq
\beq
\label{eq:18.1}
\mu_{\gamma}(x,k)=1 -\int\limits_{\RR^d} \frac{e^{i\xi x}
H_{\gamma}(k,-\xi)}{\xi^2+2(k+ i0\gamma)\xi } d\xi, \ \ k\in\RR^d\backslash0, \ \ 
\gamma\in S^{d-1},
\eeq
\beq
\label{eq:19}
\mu(x,k)=1 -\int\limits_{\RR^d} \frac{e^{i\xi x}H(k,-\xi)}{\xi^2+2k\xi } d\xi,
\ \ k\in\CC^d, \ \ \Im k\ne 0,
\eeq
where $\psi^+= e^{ikx}\mu^+$ are the eigenfunctions specified by (\ref{eq:3a}), 
(\ref{eq:3b}), $\psi= e^{ikx}\mu$  are the eigenfunctions specified by (\ref{eq:4}), 
$\mu_{\gamma}(x,k)=\mu(x,k + i 0 \gamma)$, $k\in\RR^d\backslash0$. 

The following formulas hold:
\beq
\label{eq:20}
f(k,l)=F(k,k-l), \ \  k,l\in\RR^d, \ \ k^2=l^2=E>0,
\eeq
\beq
\label{eq:20.1}
h_{\gamma}(k,l)=H_{\gamma}(k,k-l), \ \  k,l\in\RR^d, \ \ k^2=l^2=E>0, \ \ 
\gamma\in S^{d-1},
\eeq
\beq
\label{eq:21}
h(k,l)=H(k,k-l), \ \  k,l\in\CC^d, \ \ \Im k = \Im l \ne 0,  \ \ k^2=l^2=E, \ \ 
\eeq
where $f$ is the classical scattering amplitude of (\ref{eq:3a}), (\ref{eq:3b}),
$h_{\gamma}$, $h$ are the Faddeev generalized scattering data of \cite{F2}.

We recall also that for regular real-valued potentials the following formulas hold
(at least outside of the singularities of the Faddeev functions in spectral 
parameter $k$):
\beq
\label{eq:23a}
\frac{\partial}{\vphantom{\overline{d}}\partial\bar{k}_j} \psi(x,k) = 
-2\pi \int\limits_{\RR^d}\xi_j 
H(k,-\xi) \psi(x,k+\xi) \delta(\xi^2+2k\xi) d\xi,
\eeq
\beq
\label{eq:23b}
\frac{\partial}{\vphantom{\overline{d}}\partial\bar{k}_j} H(k,p) = 
-2\pi \int\limits_{\RR^d}\xi_j 
H(k,-\xi) H(k+\xi,p+\xi) \delta(\xi^2+2k\xi) d\xi,
\eeq
$j=1,\ldots,d$,\ \ $k\in\CC^d\backslash\RR^d$,\ \ $x,p\in\RR^d$,
\beq
\label{eq:23c}
\psi_{\gamma}(x,k) = \psi^{+}(x,k) + 2\pi i \int\limits_{\RR^d} h_{\gamma}(k,\xi) 
\theta((\xi-k)\gamma) \delta(\xi^2-k^2) \psi^{+}(x,\xi) d\xi,
\eeq
\beq
\label{eq:23d}
h_{\gamma}(k,l) = f(k,l) + 2\pi i \int\limits_{\RR^d} h_{\gamma}(k,\xi) 
\theta((\xi-k)\gamma) \delta(\xi^2-k^2) f(\xi,l) d\xi,
\eeq
$\gamma\in S^{d-1}$, \ \ $x,k,l\in\RR^d$, \ \ $k^2=l^2$, \\
where $\delta(t)$ is the Dirac $\delta$-function, $\theta(t)$ is the Heaviside 
step function;
\beq
\label{eq:23e}
\mu(x,k)\rightarrow 1 \ \ \mbox{for} \ \ |k|\rightarrow\infty, \ \ x\in\RR^d,
\eeq
\beq
\label{eq:23f}
H(k,p)\rightarrow \frac{1}{(2\pi)^d} \int\limits_{\RR^d} v(x) e^{ipx} dx  \ \ \mbox{for} \ \ |k|\rightarrow\infty, \ \ p\in\RR^d,
\eeq
$$
|k|=\sqrt{|\Re k|^2+|\Im k|^2},
$$
see \cite{F2}, \cite{BC}, \cite{NKh} and references therein.

Let us define the following varieties: 
\beq
\label{eq:24a}
\Sigma_E=\{k\in\CC^d:\ k^2=E \},
\eeq 
\beq
\label{eq:24b}
\Omega_{E,p}=\{ k\in\Sigma_E: \ 2kp=p^2 \}, \ \ \left\{
\begin{array}{lcc}
p=0  & \mbox{for} & d=2, \\
p\in\RR^3 &  \mbox{for} & d=3,
\end{array}\right.
\eeq
\beq
\label{eq:24c}
\Omega_{E}=\{ k\in\Sigma_E, \ \ p\in\RR^d: \ 2kp=p^2 \},
\eeq
\beq
\label{eq:24d}
\Theta_{E}=\{ k,l\in\CC^d: \ \ \Im k = \Im l, \ \ k^2=l^2=E\}.
\eeq

Note that in the present article we consider the Faddeev functions $\psi$, $H$, $h$ and 
$\psi_{\gamma}$, $H_{\gamma}$, $h_{\gamma}$ for multipoint potentials for fixed 
real energies $E$ only, for simplicity. In this connection we consider 
$$
\psi \ \ \mbox{on}  \ \ \RR^d\times(\Sigma_E\backslash\Re\Sigma_E),  \ \ 
H \ \ \mbox{on} \ \  \Omega_{E}\backslash\Re\Omega_{E}, \ \ 
h \ \ \mbox{on} \ \  \Theta_{E}\backslash\Re\Theta_{E},
$$
$$
\psi_{\gamma}(x,k), \ \ H_{\gamma}(k,p), \ \ 
h_{\gamma}(k,l) \ \ \mbox{for}
$$
$$
\gamma\in S^{d-1}, \ \ x,k,p,l\in\RR^{d}, \ \ p^2=2kp, \ \ k^2=l^2=E, \ \ 
k\gamma=0.
$$
In addition, we also consider the forms
$$
\bar\partial_k\psi = \sum\limits_{j=1}^d \frac{\partial}{\vphantom{\overline{d}}\partial\bar{k}_j} \psi(x,k) d\bar k_j,
\ \ 
\bar\partial_kH = \sum\limits_{j=1}^d \frac{\partial}{\vphantom{\overline{d}}\partial\bar{k}_j} H(k,p) d\bar k_j,
$$
on the varieties $\Sigma_E$, $\Omega_{E,p}$, respectively, 
where the $\partial/\partial\bar k_j$ derivatives of $\mu$, $H$ are given by 
(\ref{eq:23a}), (\ref{eq:23b}).

In addition, we recall that formulas (\ref{eq:23a})-(\ref{eq:23f}) give a basis
for monochromatic inverse scattering for regular potentials in two and three
dimensions, see \cite{BC}, \cite{G}, \cite{GM}, \cite{GN}, \cite{NKh}, \cite{N0},
\cite{N}, \cite{N2}.

\section{Main results} 
\label{sect:3}
By analogy with \cite{BF} we understand the multipoint potentials 
$v(x)$ from (\ref{eq:2}) as a limit for $N\rightarrow+\infty$ of non-local 
potentials 
\beq
\label{eq:6a}
V_N(x,x')=
\sum\limits_{j=1}^n\varepsilon_j(N) u_{j,N}(x) u_{j,N}(x'), 
\eeq
where
\beq
\label{eq:7}
(V_N\circ \mu)(x)= \sum\limits_{j=1}^n\varepsilon_j(N) 
\int\limits_{\RR^d} u_{j,N}(x) u_{j,N}(x') \mu(x') dx',
\eeq
\beq
\label{eq:8}
u_{j,N}(x) =\frac{1}{(2\pi)^d} \int\limits_{\RR^d} 
\hat u_{j,N}(\xi)e^{i\xi x}  d\xi, \ \ 
\hat u_{j,N}(\xi)=\left\{\begin{aligned} & e^{-i\xi z_j} & |\xi|\le N, \\ &0 &   |\xi|> N,
\end{aligned}  \right.
\eeq
$x,x',z_j\in\RR^d$, $z_m\ne z_j$ for $m\ne j$, $\varepsilon_j(N)$ are normalizing constant 
specified by (\ref{eq:3:1a}) for $d=3$ and (\ref{eq:3:1b}) for $d=2$.
It is clear that
$$
u_{j,N}(x) = u_{0,N}(x-z_j),\ \ \mbox{where} \ \ \hat 
u_{0,N}(\xi)=\left\{\begin{aligned} & 1 & |\xi|\le N, \\ &0 &   |\xi|> N.
\end{aligned}  \right.
$$

For $v=V_N$ equation (\ref{eq:6}) has the following explicit Faddeev solutions:
\beq
\label{eq:9}
\mu_N(x,k)= 1+ \frac{1}{(2\pi)^d} \int\limits_{\RR^d} 
\tilde\mu_N(\xi,k)e^{i\xi x}  d\xi, 
\eeq
\beq
\label{eq:10}
\tilde\mu_N(\xi,k)=-\frac{\sum\limits_{j=1}^n c_{j,N}(k)\hat u_{j,N}(\xi)}
{\xi^2+2k\xi},
\eeq
$x\in\RR^d$,  $\xi\in\RR^d$,  $k\in\CC^d$, $\Im k\ne0$, where 
$c_N(k)=(c_{1,N}(k),\ldots,c_{n,N}(k))$ is the solution of the following linear equation:
\beq
\label{eq:10a}
A_N(k) c_N(k) = b_N,
\eeq
where $A_N(k)$ is the $n\times n$ matrix and $b_N$ 
is the $n$-component vector with the following elements:
\beq
\label{eq:10b}
A_{m,j,N}(k)=\delta_{m,j} + \varepsilon_m(N)\frac{1}{(2\pi)^d}\int\limits_{\RR^d} 
\frac{\hat u_{m,N}(-\xi) \hat u_{j,N}(\xi)}{\xi^2+2k\xi} d\xi,
\eeq
\beq
\label{eq:10c}
b_{m,N} = \varepsilon_{m}(N).
\eeq

In addition, equation (\ref{eq:6}) has  
the following classical scattering solutions:
\beq
\label{eq:11}
\mu^+_N(x,k)= \mu_N(x,k+i0k), \ \ x\in\RR^d, \ \ k\in\RR^d\backslash0, 
\eeq
arising from 
\beq
\label{eq:12}
\tilde\mu^+_N(\xi,k)=\tilde\mu_N(\xi,k+i0k), \ \ 
\xi\in\RR^d, \ \ k\in\RR^d\backslash0.
\eeq

Let us consider the following Green functions for the operator $\Delta + 2ik\nabla$: 
\beq
\label{eq:3:4}
 g(x,k)=-\frac{1}{(2\pi)^d} \int\limits_{\RR^d} 
\frac{e^{i\xi x}}{\xi^2+2k\xi}  d\xi,\ \ x\in\RR^d \ \ k\in\CC^d, \ \ \Im k\ne 0,
\eeq
\beq
\label{eq:3.9}
 g_{\gamma}(x,k)=-\frac{1}{(2\pi)^d}\int\limits_{\RR^d} 
\frac{e^{i\xi x}}{\xi^2+2(k + i0\gamma) \xi}  d\xi, \ \ x\in\RR^d \ \ k\in\RR^d\backslash 0,
\ \ \gamma\in S^{d-1},
\eeq
\beq
\label{eq:3:15}
 g^+(x,k)=-\frac{1}{(2\pi)^d}\int\limits_{\RR^d} 
\frac{e^{i\xi x}}{\xi^2+2(k+i0k) \xi}  d\xi, \ \ x\in\RR^d \ \  k\in\RR^d\backslash0. 
\eeq
One can see that $G(x,k)=e^{ikx}g(x,k)$, where $G(x,k)$ was defined by (\ref{eq:1.6}).
Note also that for $d=3$ the Green function $g^+(x,k)$ can be calculated explicitly:
\beq
\label{eq:3:15b}
 g^+(x,k)=-\frac{1}{4\pi}\frac{e^{-ikx} e^{i|k||x|}}{|x|}. 
\eeq

\begin{theorem}
\label{th:3.1}
Let d=2, 3,
\beq
\label{eq:3:1a}
\varepsilon_j(N)= \alpha_j \left(
1-  \frac{\alpha_j N}{2\pi^2} \right)^{-1},
\ \ \alpha_j\in\RR, \ \ j=1,\ldots,n,  \ \ \mbox{for} \ \ d=3,
\eeq
\beq
\label{eq:3:1b}
\varepsilon_j(N)=
\alpha_j\left(1-\frac{\alpha_j}{2\pi}\ln(N)\right)^{-1}, \ \ \alpha_j\in\RR, 
\ \ j=1,\ldots,n,  \ \ \mbox{for} \ \ d=2, 
\eeq
Then: 
\begin{enumerate} 
\item The limiting eigenfunctions
\beq
\label{eq:3:2}
\psi(x,k)=e^{ikx} \lim\limits_{N\rightarrow+\infty} \mu_N(x,k), \ \ 
x\in\RR^d, \ \ k\in\CC^d\backslash\RR^d, \ \ k^2 = E \in\RR,
\eeq
are well-defined (at least outside the spectral singularities).
\item  The following formulas hold: 
\beq
\label{eq:3:3}
\psi(x,k)=e^{ikx}\left[
1 +  \sum\limits_{j=1}^n c_j(k) g(x-z_j,k)\right], \ \  
k\in\CC^d\backslash\RR^d, \ \ k^2 = E \in\RR,
\eeq

where 
$c(k)=(c_{1}(k),\ldots,c_{n}(k))$ is the solution of the following linear equation:
\beq
\label{eq:3:5}
\tilde A(k) c(k) = \tilde b(k),
\eeq
where $\tilde A(k)$ is the $n\times n$ matrix, $\tilde b(k)$ is the $n$-component vector 
with the following elements for $d=3$:
\begin{align}
\label{eq:3:6a}
&\tilde A_{m,j}(k) =\left\{ \begin{array}{ll} 1, & m=j \\
-\alpha_m \left( 1- \frac{\alpha_m}{4\pi}|\Im k|\right)^{-1} g(z_m-z_j,k), \ \ \  & m\ne j,
\end{array}
\right. \\ 
\label{eq:3:7a}
&\tilde b_m(k) =\alpha_m \left( 1- \frac{\alpha_m}{4\pi}|\Im k|\right)^{-1};
\end{align}
and with the following elements for $d=2$:
\begin{align}
\label{eq:3:6b}
&\tilde A_{m,j}(k) =\left\{ \begin{array}{ll} 1, & m=j \\
-\alpha_m \left( 1- \frac{\alpha_m}{2\pi}(\ln (|\Re k| + |\Im k|)  \right)^{-1} g(z_m-z_j,k), \ \ \  & m\ne j,
\end{array}
\right. \\ 
\label{eq:3:7b}
&\tilde b_m(k) =\alpha_m \left( 1- \frac{\alpha_m}{2\pi}(\ln (|\Re k| + |\Im k|) \right)^{-1}.
\end{align}
In addition, for limiting values of $\psi$ the following formulas hold:
\begin{eqnarray}
\label{eq:3:8}
\psi_{\gamma}(x,k)=\psi(x,k+ i0\gamma )= e^{ikx}\left[
1 + \sum\limits_{j=1}^n c_{\gamma,j}(k) g_{\gamma}(x-z_j,k) \right],  \\
x\in\RR^d, \ \ k\in\RR^d\backslash 0, \ \ \gamma\in S^{d-1}, \ \ k\gamma=0,
\nonumber
\end{eqnarray}

where 
$c_{\gamma}(k)=(c_{\gamma,1}(k),\ldots,c_{\gamma,n}(k))$ is the solution of the following 
linear equation:
\beq
\label{eq:3:10}
\tilde A_\gamma(k) c_{\gamma}(k) = \tilde b_{\gamma}(k),
\eeq
where 
\beq
\label{eq:3:11}
\tilde A_\gamma(k)=\tilde A(k+i0\gamma), \ \ \tilde b_{\gamma}(k)=\tilde b(k+i0\gamma).
\eeq
\item The Faddeev generalized scattering data for the limiting potential 
$v=\lim\limits_{N\rightarrow+\infty} V_N$, associated with the limiting 
eigenfunctions  $\psi$,  $\psi_{\gamma}$, are given by:
\begin{eqnarray}
\label{eq:3:12}
h(k,l) = \frac{1}{(2\pi)^d}  \sum\limits_{j=1}^n c_{j}(k) e^{i(k-l)z_j},\\
k,l\in\CC^3, \ \ \Im k=\Im l \ne0, \ \   k^2=l^2=E\in\RR\nonumber,
\end{eqnarray}
where $c_{j}(k)$ are the same as in (\ref{eq:3:3}),  (\ref{eq:3:5});
\begin{eqnarray}
\label{eq:3:13}
h_{\gamma}(k,l) = \frac{1}{(2\pi)^d}  \sum\limits_{j=1}^n c_{\gamma,j}(k) e^{i(k-l)z_j},\\
k,l\in\RR^d\backslash 0, \ \ k^2=l^2=E, \ \ \gamma\in S^{d-1}, \ \ k\gamma=0, \nonumber
\end{eqnarray}
where $c_{\gamma,j}(k)$ are the same as in (\ref{eq:3:8}),  (\ref{eq:3:10}).
\end{enumerate}
\end{theorem}

Note that if $\|\tilde b(k)\|=\infty$ then we understand (\ref{eq:3:3})-(\ref{eq:3:11}) as
(\ref{eq:4.9}), (\ref{eq:4.11})-(\ref{eq:4.13}), (\ref{eq:4.23}), (\ref{eq:4.25})-(\ref{eq:4.27}).

\begin{remark}
Let the assumptions of Theorem~\ref{th:3.1} be fulfilled. Then:
\begin{enumerate}
\item
For the classical scattering eigenfunctions $\psi^+$  the following formulas hold:
\beq
\label{eq:3:14}
\psi^+(x,k)=e^{ikx}\left[
1 +  \sum\limits_{j=1}^n c^+_j(k) g^+(x-z_j,k)\right],
\eeq
where 
$c^+(k)=(c^+_{1}(k),\ldots,c^+_{n}(k))$ is the solution of the following linear equation:
\beq
\label{eq:3:16}
\tilde A^+(k) c^+(k) = \tilde b^+(k),
\eeq
where $\tilde A^+(k)$ is the $n\times n$ matrix, and $\tilde b^+(k)$ is the $n$-component 
vector with the following elements for $d=3$:
\beq
\label{eq:3:16a}
\tilde A^+_{m,j}(k)=\left\{ \begin{array}{ll} 1 & m=j \\
-\alpha_m \left( 1+ \frac{i \alpha_m}{4\pi}|k|\right)^{-1} g^+(z_m-z_j,k), \ \ \  & m\ne j,
\end{array}
\right.
\eeq
\beq
\label{eq:3:16b}
\tilde b^+_m(k) =\alpha_m \left( 1+ \frac{i\alpha_m}{4\pi}|k|\right)^{-1};
\eeq
and  with the following elements for $d=2$:
\beq
\label{eq:3:16c}
\tilde A^+_{m,j}(k)=\left\{ \begin{array}{ll} 1 & m=j \\
-\alpha_m \left( 1+\frac{ \alpha_m}{4\pi}(\pi i -2\ln |k|)\right)^{-1} g^+(z_m-z_j,k), \ \ \  & m\ne j,
\end{array}
\right.
\eeq
\beq
\label{eq:3:18a}
\tilde b^+_m(k) =\alpha_m \left( 1+ \frac{\alpha_m}{4\pi}(\pi i -2 \ln |k|)\right)^{-1};
\eeq
\item 
For the classical scattering amplitude $f$ the following formula holds:
\begin{eqnarray}
f(k,l) = \frac{1}{(2\pi)^d}  \sum\limits_{j=1}^n c^+_{j}(k) e^{i(k-l)z_j}, 
\label{eq:3:19}\\
k,l\in\RR^d, \ \ k^2=l^2=E\in\RR\nonumber,
\end{eqnarray}
where $c^+_{j}(k)$ are the same as in (\ref{eq:3:14}),  (\ref{eq:3:16}). 
In a slightly different form formulas (\ref{eq:3:14}) -  (\ref{eq:3:19}) are contained
in Section~II.1.5 and Chapter~II.4 of \cite{AGHH}. In addition, the classical 
scattering functions  $\psi^+$ and $f$ for $d=3$ are expressed in terms of 
elementary functions via (\ref{eq:3:14})- (\ref{eq:3:19}).
\end{enumerate}
\end{remark}

\begin{proposition}
\label{pr:3.1}
Formulas  (\ref{eq:23a}),(\ref{eq:23b}) in terms of $\bar\partial_k\mu$, 
$\bar\partial_k H$, on $\Sigma_E$, $\Omega_{E,p}$, formulas (\ref{eq:23c}), 
(\ref{eq:23d}) with 
$k\gamma=0$ and formula (\ref{eq:23e}) for $|\Im k|\rightarrow\infty$  are fulfilled 
for functions $\psi=e^{ikx}\mu $, $\psi_{\gamma}$,  $\psi^+$, $h$, $h_{\gamma}$ 
of Theorem~\ref{th:3.1}, at least for $x\ne z_j$, $j=1,\ldots,n$.
\end{proposition}

\begin{statement}
\label{st:3.1}
Let $d=3$, $n=2$, $E=E_{\mbox{\scriptsize fix}}>0$. Then for appropriate 
$\alpha_1, \ \alpha_2\in\RR\backslash0$, $z_1,z_2\in\RR^3$ there are real spectral singularities 
$k=k'+i0\gamma'$ with $\gamma'\in S^2$, $k'\in\RR^3$, $(k')^2=E_{\mbox{\scriptsize fix}}$, 
$k'\gamma'=0$, of the Faddeev functions $\psi$, $h$ of Theorem~\ref{th:3.1}.
\end{statement}
\begin{remark}
\label{rem:3.1}
In connection with Statement~\ref{st:3.1}, note that for the case $d=3$, $n=1$, 
studied in the old unpublished work of Faddeev, there are no real spectral 
singularities of the Faddeev functions $\psi$, $h$. In addition, in \cite{GRN}
it was shown that for the case $d=2$, $n=1$, $\alpha\in\RR\backslash0$ the 
Faddeev functions always have some real spectral singularities 
(see Statement~3.1 of \cite{GRN} for details).
\end{remark}

Let us recall that $\dim_{\CC}\Sigma_E=1$, $ \dim_{\RR}\Sigma_E=2$ for $d=2$. In addition,
it is known that for a fixed real energy $E=E_{\mbox{\scriptsize fix}}$  the spectral singularities of $\psi$ and $H$
on $\Sigma_E\backslash\Re\Sigma_E$ are zeroes of a real-valued determinant function (for real potentials). 
Thus, one can expect that these spectral singularities on $\Sigma_{E_{\mbox{\tiny fix}} }$ for generic real 
potentials are either empty or form a family of curves $\Gamma_j$, $j=\pm1,\pm2,\ldots\pm{J}$ . 
The problem of studying the geometry of these  spectral singularities on $\Sigma_{E_{\mbox{\tiny fix}}}$ 
was formulated already in \cite{GN}. In addition, it was expected in \cite{GN} that the most natural 
configuration of curves is a ``nest''
\beq
\label{eq:3:20}
[\Gamma_{-J}\subset\Gamma_{-J+1}\subset\ldots\subset\Gamma_{-1}
\subset S^1\subset\Gamma_{1}\subset\ldots\subset\Gamma_{J}],
\eeq
see  \cite{GN} for details. 

Figures Fig.~1--Fig.~4 show these spectral singularities for 2-point 
potentials for some interesting cases. These figures show that the geometry of the singular curves $\Gamma_j$ may be different from the ``nest''. 

\begin{center}
\parbox{5cm}{\begin{center}
\boxed{\epsfxsize=5cm \epsffile{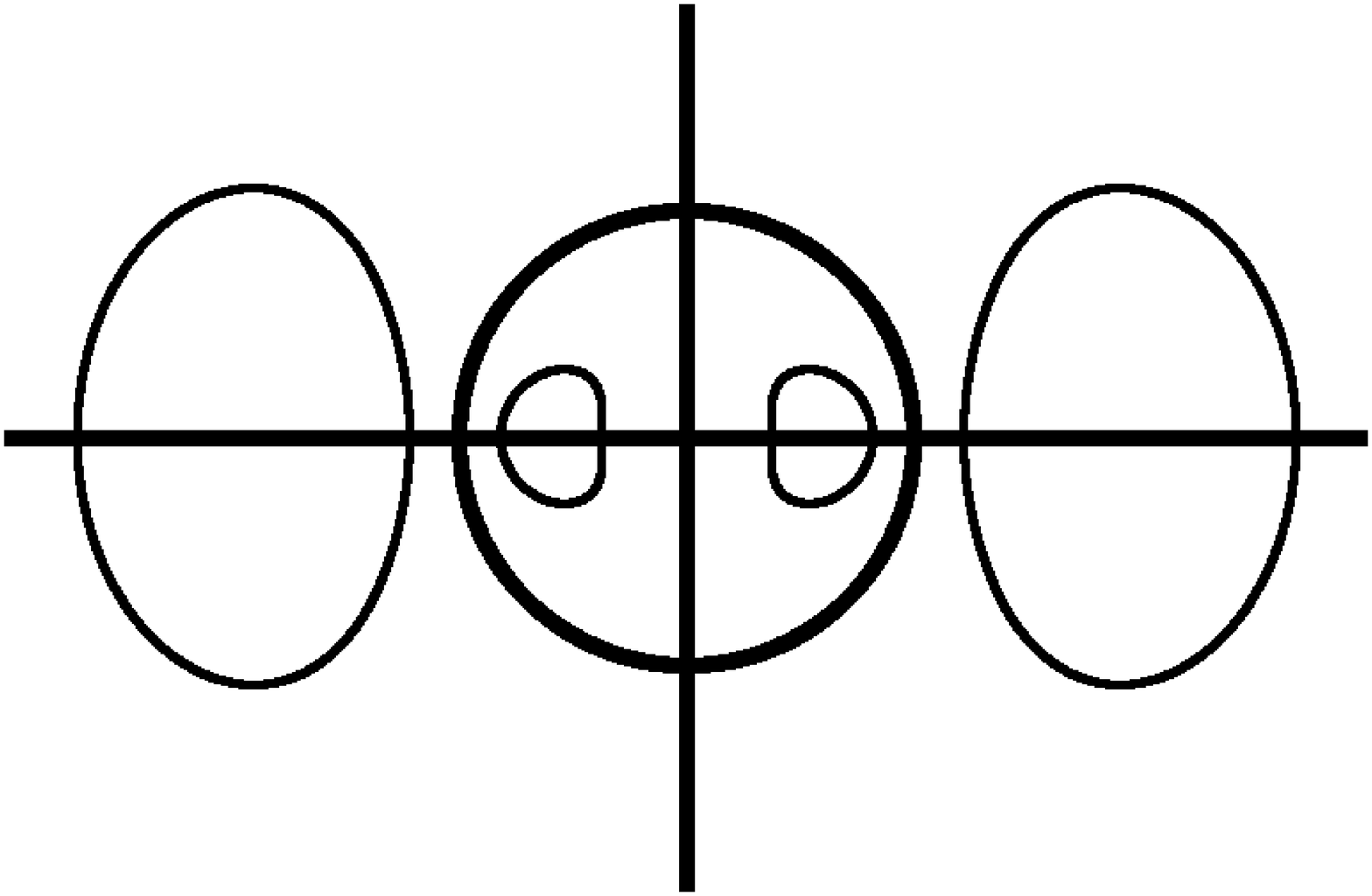}}\\ Fig. 1 \\
$E=4, \ \ z_2-z_1 = (0.5\, ,\, 0)$,  \\ $\alpha_1 =5, \ \ \alpha_2 = 6$   
\end{center}}    \hspace{5mm}  
\parbox{5cm}{\begin{center}
\boxed{\epsfxsize=5cm \epsffile{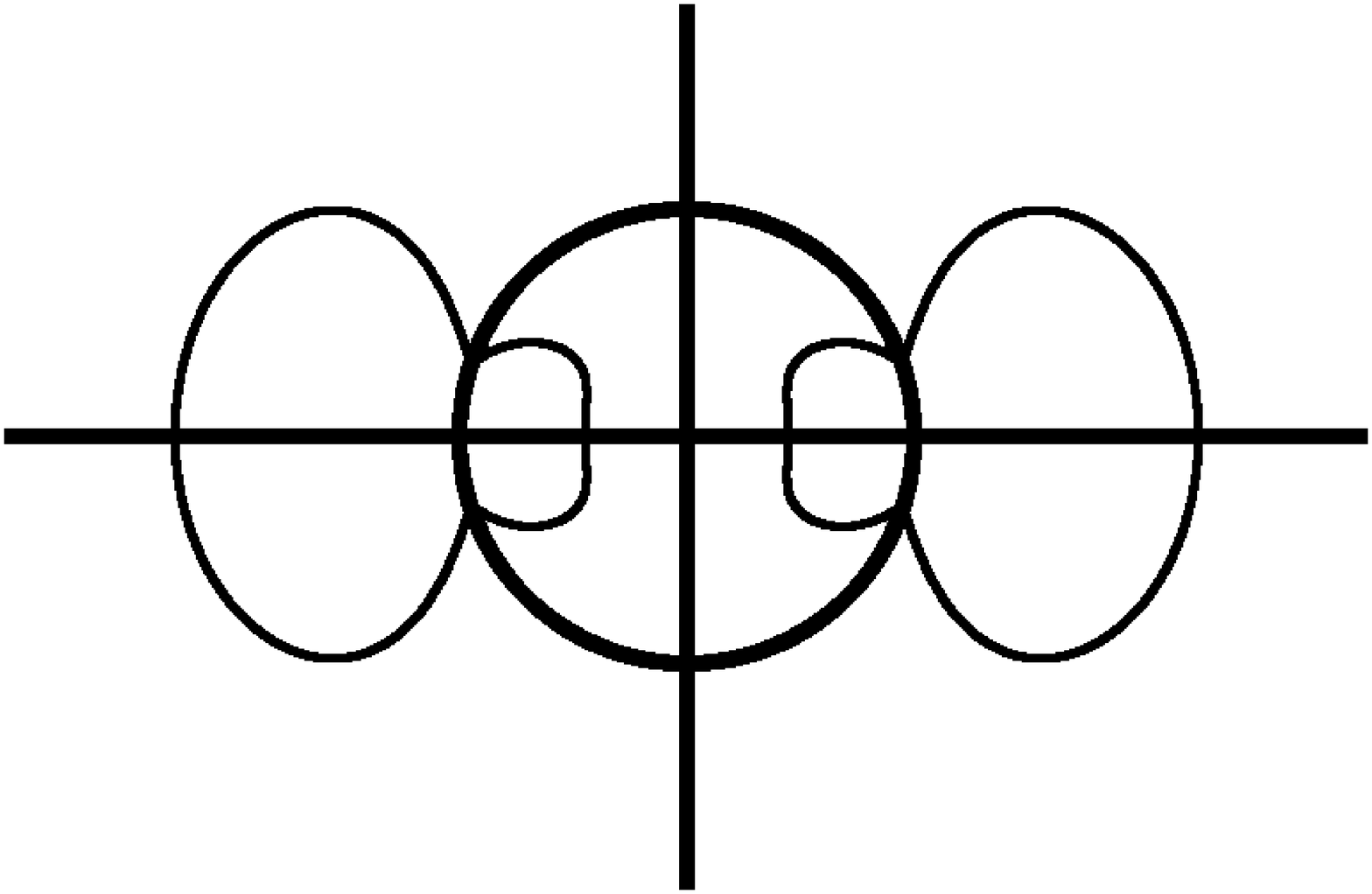}}\\ Fig. 2  \\ 
$E=6, \ \ z_2-z_1 = (0.5\, ,\, 0)$,  \\ $\alpha_1 =5, \ \ \alpha_2 = 6$  
\end{center}} 
\end{center}

\begin{center}
\parbox{5cm}{\begin{center}
\boxed{\epsfxsize=5cm \epsffile{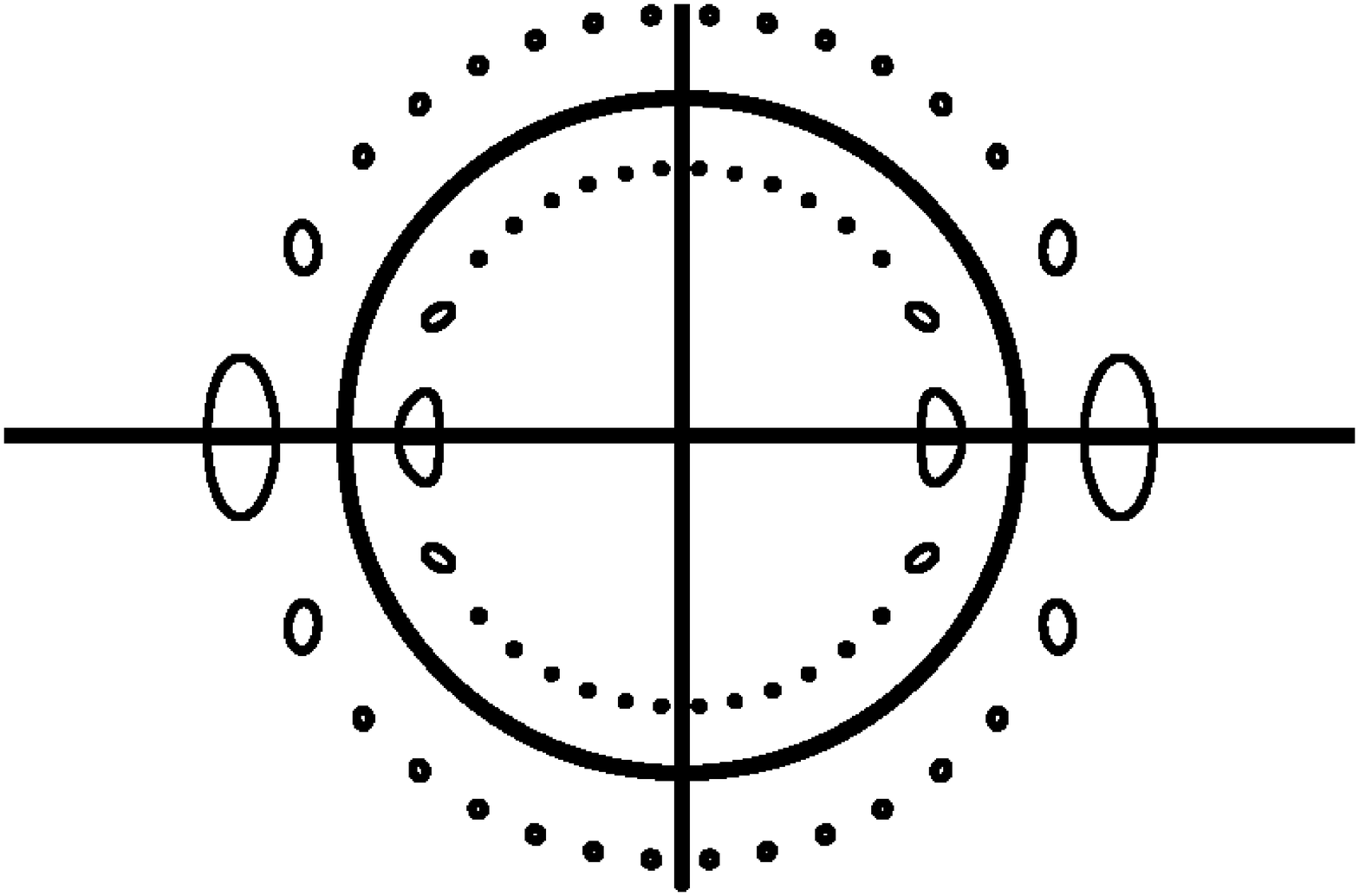}}\\ Fig. 3  \\
$E=5, \ \ z_2-z_1 = (10\, ,\, 0)$,  \\ $\alpha_1 =6, \ \ \alpha_2 = 6$   
\end{center}}    \hspace{5mm}  
\parbox{5cm}{\begin{center}
\boxed{\epsfxsize=5cm \epsffile{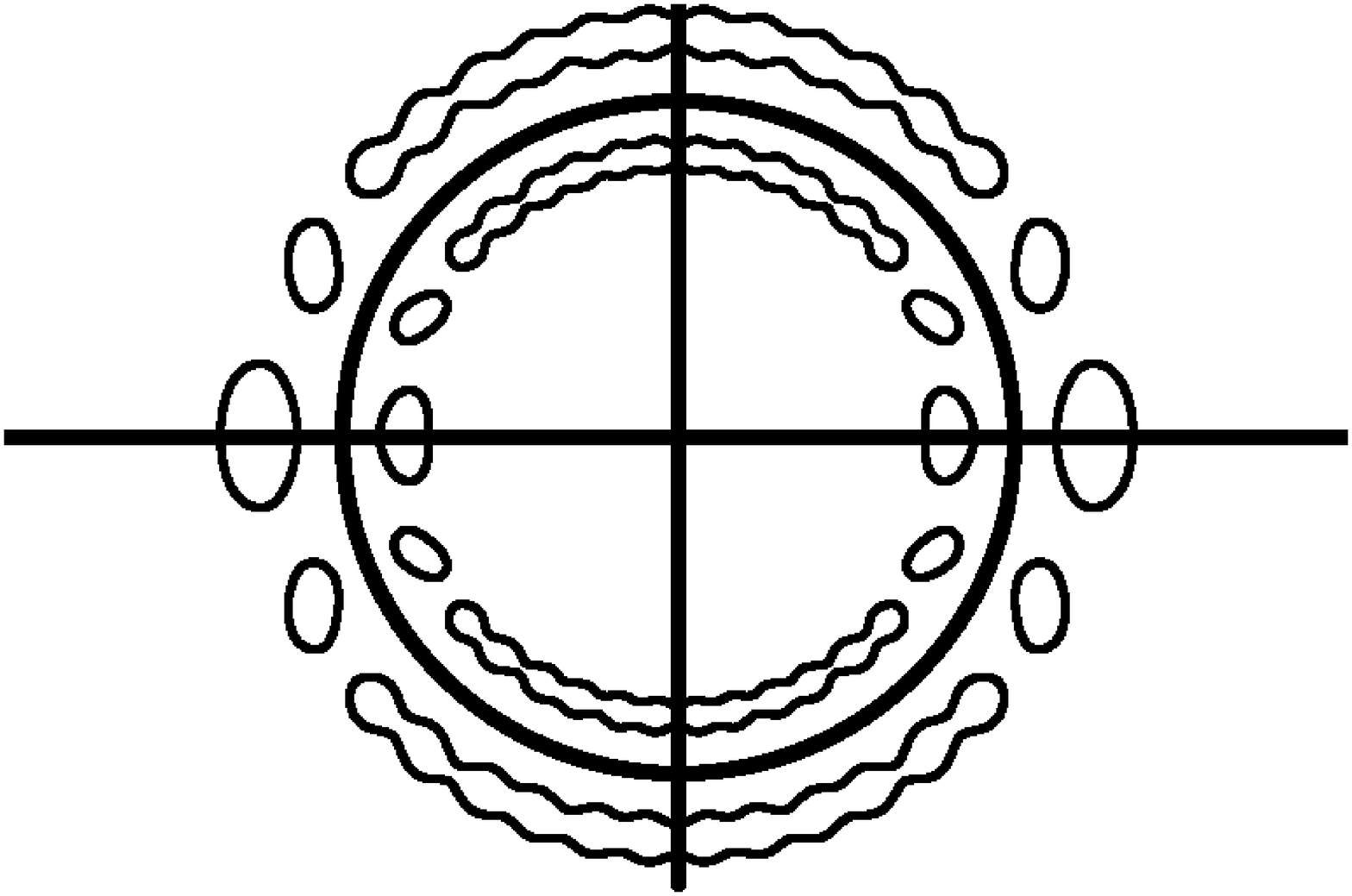}}\\ Fig. 4  \\ 
$E=5, \ \ z_2-z_1 = (10\, ,\, 0)$,  \\ $\alpha_1 =6, \ \ \alpha_2 = 6.8$  
\end{center}} 
\end{center}

In Figures~1-4 the surface $\Sigma_E$ is shown as $\CC\backslash0$ with the coordinate  
$\lambda$, where the parametrization of $\Sigma_E$ is given by the formulas:
\beq
\label{eq:3:21}
k_1= \left(\frac{1}{\lambda}
+\lambda \right)\frac{\sqrt{E}}{2},   \ \  k_2=\left(\frac{1}{\lambda}
-\lambda \right)\frac{i\sqrt{E}}{2}, \ \ \ \ \ \lambda\in\CC\backslash0.
\eeq
The coordinate axes $\Im\lambda=0$, $\Re\lambda=0$  and the unit circle  $|\lambda|=1$ 
in $\CC$ are shown in bold. This unit circle corresponds to $\Sigma_E\cap\RR^2$, i.e.
to real (physical) momenta $k=(k_1,k_2)$. The other black sets inside the rectangles 
in Figures~1-4 show singular curves $\Gamma_j$.  

\section{Sketch of proofs} 

To prove Theorem~\ref{th:3.1} we proceed from formulas (\ref{eq:8})-(\ref{eq:10c}). 
We rewrite (\ref{eq:10a}) as

\beq
\label{eq:4.1}
\left(I+\Lambda_N^{-1}(k)\, B_N(k) \right) c_N(k) = \Lambda_N^{-1}(k)\, b_N,
\eeq
where $\Lambda_N(k)$ and $B_N(k)$ are the diagonal and off-diagonal parts of
$A_N(k)$, respectively. One can see that 
\beq
\label{eq:4.2}
(\Lambda_N^{-1}(k)\,b_N)_m = \frac{\varepsilon_m(N)}
{1+ \varepsilon_m(N)\, \frac{1}{(2\pi)^d}\int_{\RR^d} 
\frac{\hat u_{m,N}(-\xi) \hat u_{m,N}(\xi)}{\xi^2+2k\xi} d\xi},
\eeq
\beq
\label{eq:4.3}
(\Lambda_N^{-1}(k)\,B_N(k))_{m,j} = (1-\delta_{m,j})\,\frac{\varepsilon_m(N)\, 
\frac{1}{(2\pi)^d}\int_{\RR^d} 
\frac{\hat u_{m,N}(-\xi) \hat u_{j,N}(\xi)}{\xi^2+2k\xi} d\xi}
{1+ \varepsilon_m(N)\, \frac{1}{(2\pi)^d}\int_{\RR^d} 
\frac{\hat u_{m,N}(-\xi) \hat u_{m,N}(\xi)}{\xi^2+2k\xi} d\xi}.
\eeq
In addition, for $N\rightarrow+\infty$:
\beq
\label{eq:4.4}
\frac{1}{(2\pi)^d}\int_{\RR^d} 
\frac{\hat u_{m,N}(-\xi) \hat u_{j,N}(\xi)}{\xi^2+2k\xi} d\xi\rightarrow -g(z_m-z_j,k), \ \ 
j\ne m, \ \ 
\mbox{for} \ \ d=2,3,
\eeq
\beq
\label{eq:4.5}
\varepsilon_m(N) \frac{1}{(2\pi)^d}\int_{\RR^d} 
\frac{\hat u_{m,N}(-\xi) \hat u_{m,N}(\xi)}{\xi^2+2k\xi} d\xi\rightarrow 
\frac{\alpha_m}{1- \frac{\alpha_m}{4\pi}|\Im k|} 
\ \ \mbox{for} \ \ d=3,
\eeq
\beq
\label{eq:4.6}
\varepsilon_m(N) \frac{1}{(2\pi)^d}\int_{\RR^d} 
\frac{\hat u_{m,N}(-\xi) \hat u_{m,N}(\xi)}{\xi^2+2k\xi} d\xi\rightarrow 
\frac{\alpha_m}{1- \frac{\alpha_m}{2\pi}(\ln (|\Re k| + |\Im k|) }
\ \ \mbox{for} \ \ d=2,
\eeq
$k\in\CC^d\backslash\RR^d$, $k^2=E\in\RR$.

One can see that (\ref{eq:4.4}) follows from (\ref{eq:3:4}) and the definition of 
$\hat u_{j,N}$ in (\ref{eq:8}). In turn, formulas (\ref{eq:4.5}),  (\ref{eq:4.6}) follow from 
(\ref{eq:3:1a}), (\ref{eq:3:1b}), the definition of $\hat u_{j,N}$ and the 
following asymptotic formulas for $N\rightarrow+\infty$:
\beq
\label{eq:4.7}
\int\limits_{\xi\in\RR^d, \ |\xi|\le N} \frac{e^{i\xi x}}{\xi^2+2 k \xi}
d\xi = 4\pi N - 2 \pi^2 |\Im k|  +O(N^{-1}) 
\ \ \mbox{for} \ \ d=3,
\eeq
\beq
\label{eq:4.8}
\int\limits_{\xi\in\RR^d, \ |\xi|\le N} \frac{e^{i\xi x}}{\xi^2+2 k \xi}
d\xi = 2\pi \ln N - 2 \pi\ln(|\Re k|+|\Im k|)+O(N^{-1})
\ \ \mbox{for} \ \ d=2,
\eeq
where $k\in\CC^d\backslash\RR^d$, $k^2=E\in\RR$.

Formulas (\ref{eq:3:2})-(\ref{eq:3:7b}) follow from (\ref{eq:8})-(\ref{eq:10}), 
(\ref{eq:4.1})-(\ref{eq:4.6}). 

Formulas (\ref{eq:3:8})-(\ref{eq:3:11}) follow from 
(\ref{eq:3:3})-(\ref{eq:3:7b}).

Formulas (\ref{eq:3:12})-(\ref{eq:3:13}) follow from the relations $\psi = e^{ikx}\mu$,
$\psi_{\gamma} = e^{ikx}\mu_{\gamma}$, and formulas (\ref{eq:18.1}), (\ref{eq:19}), 
 (\ref{eq:20.1}), (\ref{eq:21}), (\ref{eq:3:4}),(\ref{eq:3.9}),  (\ref{eq:3:3}), 
(\ref{eq:3:8}).

This completes the sketch of proof of Theorem~\ref{th:3.1}.

To prove Proposition~\ref{pr:3.1} we rewrite (\ref{eq:3:3})-(\ref{eq:3:7b}),  
(\ref{eq:3:12}) in the following form:
\beq
\label{eq:4.9}
\psi(x,k)=e^{ikx} +  \sum\limits_{j=1}^n {\cC}_j(k) G(x-z_j,k),
\eeq
\beq
\label{eq:4.10}
H(k,p) = \frac{1}{(2\pi)^d}  \sum\limits_{j=1}^n {\cC}_{j}(k) e^{-ik z_j} e^{ip z_j},\
\eeq
\beq
\label{eq:4.11}
\cA\, \cC =  \cB,
\eeq
\begin{align}
&\cA_{m,m}(k) &= &\ \ \alpha_m^{-1} - (4\pi)^{-1} |\Im k|, \ \ &d=3,\nonumber\\ 
\label{eq:4.12}
&\cA_{m,m}(k) &= &\ \ \alpha_m^{-1} - (2\pi)^{-1} \ln(|\Re k|+|\Im k|), \ \ &d=2,\\
&\cA_{m,j}(k) &= &\ \ -G(z_m-z_j,k), \ \ &m\ne j,\nonumber
\end{align}
\beq
\label{eq:4.13}
\cB_m(k) = e^{ikz_m},
\eeq
where $k\in\CC^d\backslash\RR^d$, $k^2 = E \in\RR$, $p\in\RR^d$, $p^2=2kp$, $G$ is defined 
by (\ref{eq:1.6}).

Here
$$
\cC_j(k) = e^{ikz_j} c_j(k).
$$

We recall the formulas (see \cite{NKh})

\beq
\label{eq:4.14}
\frac{\partial}{\vphantom{\overline{d}}\partial\bar{k}_j} G(x,k) = 
-\frac{1}{(2\pi)^{d-1}} \int\limits_{\RR^d}\xi_j 
e^{i(k+\xi)x} \delta(\xi^2+2k\xi) d\xi, \ \ j=1,\ldots,d.
\eeq
\beq
\label{eq:4.15}
 G(x,k+\xi)=G(x,k), \ \ \mbox{for} \ \ \xi\in\RR^d, \ \ \xi^2+2k\xi=0,
\eeq
where $k\in\CC^d\backslash\RR^d$.

We will use also the following formula:
\beq
\label{eq:4.16}
\bar\partial_k \cA_{m,m}(k) = \frac{1}{(2\pi)^{d-1}} \int\limits_{\RR^d}
\left(\sum\limits_{j=1}^d \xi_j d \bar k_j\right) \,
\delta(\xi^2+2k\xi)\, d\xi \ \ \mbox{on} \ \ \Sigma_E\backslash\Re\Sigma_E, 
\ \ E\in\RR.
\eeq

The proof of the $\bar\partial$-equation (\ref{eq:23a}) for   
$\bar\partial_k \psi(x,k)$ on $\Sigma_E\backslash\Re\Sigma_E$ 
can be sketched as formulas (\ref{eq:4.17})-(\ref{eq:4.22}) 
on $\Sigma_E\backslash\Re\Sigma_E$ as follows.

We have
\beq
\label{eq:4.17}
\bar\partial_k \psi(x,k)=  \sum\limits_{j=1}^n {\cC}_j(k) (\bar\partial_k G(x-z_j,k))+
 \sum\limits_{j=1}^n  (\bar\partial_k{\cC}_j(k)) G(x-z_j,k).
\eeq
Using (\ref{eq:4.10}), (\ref{eq:4.14}) one can see that:
\beq
\label{eq:4.18}
\sum\limits_{j=1}^n {\cC}_j(k) (\bar\partial_k G(x-z_j,k))=
-2\pi \int\limits_{\RR^d}\left(\sum\limits_{s=1}^d \xi_s d\bar k_s\right)
H(k,-\xi)e^{i(k+\xi)x} \delta(\xi^2+2k\xi) d\xi.
\eeq
Taking into account (\ref{eq:4.9}), (\ref{eq:4.10}), (\ref{eq:4.17}), (\ref{eq:4.18})
one can see that to prove equation (\ref{eq:23a}) it is sufficient to verify the following
$\bar\partial$ equation:
\beq
\label{eq:4.19}
\bar\partial_k \cC_m(k)= -(2\pi)^{d-1} \int\limits_{\RR^d} 
\left(\sum\limits_{s=1}^d \xi_s d\bar k_s\right)\left[ \sum\limits_{j=1}^n
\cC_j(k) e^{-i(k+\xi)z_j} \cC_j(k+\xi)\right] \delta(\xi^2+2k\xi) d\xi.
\eeq
In turn, (\ref{eq:4.19}) follows form the following formulas:
\beq
\label{eq:4.20}
(\bar\partial_k \cC)\, \cA+  \cC\, (\bar\partial_k \cA) =0,  
\eeq
\beq
\label{eq:4.21}
\bar\partial_k \cA_{m,j}(k) =  \frac{1}{(2\pi)^{d-1}} \int\limits_{\RR^d}
\left(\sum\limits_{s=1}^d \xi_s d\bar k_s\right)
e^{i(k+\xi)z_m}e^{-i(k+\xi)z_j}  \delta(\xi^2+2k\xi) d\xi,
\eeq
\beq
\label{eq:4.22}
(\cA^{-1}\bar\partial_k \cA)_{m,j}(k) = \frac{1}{(2\pi)^{d-1}} \int\limits_{\RR^d}
\left(\sum\limits_{s=1}^d \xi_s d\bar k_s\right)
\cC_m(k+\xi)e^{-i(k+\xi)z_j}  \delta(\xi^2+2k\xi) d\xi.
\eeq

The $\bar\partial$-equation  (\ref{eq:23b}) for  $\bar\partial_k H$ on 
$\Sigma_E\backslash\Re\Sigma_E$ follows from formula (\ref{eq:19}) and the 
$\bar\partial$-equation (\ref{eq:23a}) for   
$\bar\partial_k \psi$ on $\Sigma_E\backslash\Re\Sigma_E$.

To verify (\ref{eq:23c}) with $k\gamma=0$ we rewrite 
(\ref{eq:3:8})-(\ref{eq:3:11}), (\ref{eq:3:13}) and  
(\ref{eq:3:14})-(\ref{eq:3:19}) in a similar way with 
(\ref{eq:4.9})-(\ref{eq:4.13}):

\beq
\label{eq:4.23}
\psi_{\gamma}(x,k)=e^{ikx} +  \sum\limits_{j=1}^n {\cC}_{\gamma,j}(k) 
G_{\gamma}(x-z_j,k),
\eeq
\beq
\label{eq:4.24}
h_{\gamma}(k,l) = \frac{1}{(2\pi)^d}  \sum\limits_{j=1}^n {\cC}_{\gamma,j}(k) 
e^{-il z_j},\
\eeq
\beq
\label{eq:4.25}
\cA_{\gamma}\, \cC_{\gamma} =  \cB_{\gamma},
\eeq
\begin{align}
&\cA_{\gamma,m,m}(k) &= &\ \ \alpha_m^{-1}, \ \ &d=3,\nonumber\\ 
\label{eq:4.26}
&\cA_{\gamma,m,m}(k) &= &\ \ \alpha_m^{-1} - (2\pi)^{-1} \ln(|k|), \ \ &d=2,\\
&\cA_{\gamma,m,j}(k) &= &\ \ -G_{\gamma}(z_m-z_j,k), \ \ &m\ne j,\nonumber
\end{align}
\beq
\label{eq:4.27}
\cB_{\gamma,m}(k) = e^{ikz_m},
\eeq
where $\gamma\in S^{d-1}$, $k,l\in\RR^d\backslash0$, $k\gamma=0$, 
$G_{\gamma}(x,k)=G(x,k+i0\gamma)$;
\beq
\label{eq:4.28}
\psi^{+}(x,k)=e^{ikx} +  \sum\limits_{j=1}^n {\cC}^{+}_{j}(k) 
G^{+}(x-z_j,k),
\eeq
\beq
\label{eq:4.29}
f(k,l) = \frac{1}{(2\pi)^d}  \sum\limits_{j=1}^n {\cC}^{+}_{j}(k) 
e^{-il z_j},\
\eeq
\beq
\label{eq:4.30}
\cA^{+}\, \cC^{+} =  \cB^{+},
\eeq
\begin{align}
&\cA^{+}_{m,m}(k) &= &\ \ \alpha_m^{-1} + i(4\pi)^{-1}|k| , \ \ &d=3,\nonumber\\ 
\label{eq:4.31}
&\cA^{+}_{m,m}(k) &= &\ \ \alpha_m^{-1} + (4\pi)^{-1}(\pi i -2 \ln(|k|)), \ \ &d=2,\\
&\cA^{+}_{m,j}(k) &= &\ \ -G^{+}(z_m-z_j,k), \ \ &m\ne j,\nonumber
\end{align}
\beq
\label{eq:4.32}
\cB^{+}_{m}(k) = e^{ikz_m},
\eeq
where $k,l\in\RR^d\backslash0$.

We recall the formula (see \cite{F2}, \cite{NKh}):
\beq
\label{eq:4.33}
G_{\gamma}(x,k) = 
G^{+}(x,k) +\frac{2\pi i}{(2\pi)^d}\int\limits_{\xi\in\RR^d} e^{i\xi x} 
\delta(\xi^2-k^2) \theta((\xi-k)\gamma)d\xi,
\eeq
where $\gamma\in S^{d-1}$, $k\in\RR^d\backslash0$.

We will use also the following formula:
\beq
\label{eq:4.34}
\cA_{\gamma,m,m}(k) = \cA^{+}_{m,m}(k) -\frac{2\pi i}{(2\pi)^d}\int\limits_{\xi\in\RR^d} \delta(\xi^2-k^2) 
\theta(\xi \gamma)d\xi,
\eeq
where $\gamma\in S^{d-1}$, $k\in\RR^d\backslash0$, $k\gamma=0$. 

One can see that for $\psi_{\gamma}$, $\psi^+$ of (\ref{eq:4.23}), (\ref{eq:4.28}) 
relation (\ref{eq:23c}) with $k\gamma=0$ is reduced to the following two relations:
\begin{align}
\label{eq:4.35}
\sum\limits_{j=1}^n \cC_{\gamma,j}(k) \left(G_{\gamma}(x-z_j,k) -  G^{+}(x-z_j,k)\right) = \\
=2\pi i\int_{\RR^d} h_\gamma(k,\xi) e^{i\xi x}\delta(\xi^2-k^2) \theta(\xi\gamma) d\xi, \nonumber 
\end{align}
\beq
\label{eq:4.36}
\cC_{\gamma,j}(k) =  \cC^{+}_{j}(k)
+2\pi i\int_{\RR^d} h_\gamma(k,\xi) \delta(\xi^2-k^2) \theta(\xi\gamma) \cC^{+}_{j}(\xi)  d\xi,
\eeq
where $\gamma\in S^{d-1}$, $k\in\RR^d\backslash0$, $k\gamma=0$. 

Relation  (\ref{eq:4.35}) follows from  (\ref{eq:4.33}) and  (\ref{eq:4.24}).  Relation 
(\ref{eq:4.36}) follows from the following relations
\beq
\label{eq:4.37}
(I+(\cA^{+})^{-1}(\cA_{\gamma}-\cA^{+})) \cC_{\gamma} = \cC^{+},
\eeq
\beq
\label{eq:4.38}
(\cA_{\gamma}(k)-\cA^{+}(k))_{m,j}=-\frac{2\pi i}{(2\pi)^d}\int\limits_{\xi\in\RR^d} e^{i\xi (z_m-z_j)} 
\delta(\xi^2-k^2) \theta(\xi \gamma)d\xi,
\eeq
\beq
\label{eq:4.39}
[(\cA^{+}(k))^{-1}(\cA_{\gamma}(k)-\cA^{+}(k))]_{m,j}=
-\frac{2\pi i}{(2\pi)^d}\int\limits_{\xi\in\RR^d} \cC^{+}_{m}(\xi)   e^{-i\xi z_j} 
\delta(\xi^2-k^2) \theta(\xi \gamma)d\xi,
\eeq
and formula (\ref{eq:4.24}) for $h_{\gamma}$.

This completes the sketch of proof of the relation (\ref{eq:23c}). 

Relation (\ref{eq:23d}) can be obtained using  (\ref{eq:18}),  
(\ref{eq:18.1}), (\ref{eq:20}),  (\ref{eq:20.1}), (\ref{eq:23c}).

Formula (\ref{eq:23e}) for $|\Im k|\rightarrow\infty$ can be obtained using 
(\ref{eq:3:3})-(\ref{eq:3:7b}).

Sketch of proof of Proposition~\ref{pr:3.1} is completed.

To prove Statement~\ref{st:3.1} we point out that spectral singularities of $\psi$, $h$ 
on $\Sigma_E$, $E\in\RR$, coincide with the zeroes of $\det\cA(k)$, where $\cA(k)$ is defined 
by (\ref{eq:4.12}) (we can always assume that all $\alpha_m\ne0$). For $d=3$, $n=2$ we have that
\beq
\label{eq:4.40}
\det\cA(k) = \left[\frac{1}{\alpha_1} - \frac{|\Im k|}{4\pi} \right] \cdot  
\left[\frac{1}{\alpha_2} - \frac{|\Im k|}{4\pi} \right] - G(z_1-z_2,k)\cdot G(z_2-z_1,k).
\eeq
We recall that $G(x,k)$ is real-valued (see \cite{NKh}) or, more precisely,
\beq
\label{eq:4.41}
G(x,k) = \overline{G(x,k)}, \ \ k\in\Sigma_E\backslash\Re\Sigma_E, \ \ E\in\RR.
\eeq
For $k=k'+i0\gamma'$ of Statement~\ref{st:3.1} formulas (\ref{eq:4.40}), (\ref{eq:4.41}) 
take the form:
\beq
\label{eq:4.42}
\det\cA(k'+i0\gamma') = \frac{1}{\alpha_1\alpha_2 } - G_{\gamma'}(z_1-z_2,k')\cdot G_{\gamma'}(z_2-z_1,k').
\eeq
\beq
\label{eq:4.43}
G_{\gamma'}(x,k') = \overline{G_{\gamma'}(x,k')}.
\eeq
Therefore, for $z_1$, $z_2$ such that $G_{\gamma'}(z_1-z_2,k')\cdot G_{\gamma'}(z_2-z_1,k')\ne 0$ 
one can always choose $\alpha_1, \ \alpha_2 \in\RR$ such that $\det\cA(k'+i0\gamma')=0$. 

Statement~\ref{st:3.1} is proved.

\end{document}